# A Single Pair of Weyl Fermions in Half-metallic EuCd$_2$As$_2$ Semimetal

*Lin-Lin Wang[1*], Na Hyun Jo[2], Brinda Kuthanazhi[2], Yun Wu[2], Robert J. McQueeney[1,2], Adam Kaminski[1,2], and Paul C. Canfield[1,2]*

[1]Ames Laboratory, U.S. Department of Energy, Ames, IA 50011, USA

[2]Department of Physics and Astronomy, Iowa State University, Ames, IA 50011, USA

# Abstract

An ideal Weyl semimetal with a single pair of Weyl points (WPs) may be generated by splitting a single Dirac point (DP) through the breaking of time-reversal symmetry by magnetic order. However, most known Dirac semimetals possess a pair of DPs along an axis that is protected by crystalline symmetry. Here, we demonstrate that a single pair of WPs may also be generated from a pair of DPs. Using first-principles band structure calculations, we show that inducing ferromagnetism in the AFM Dirac semimetal $EuCd_2As_2$ generates a single pair of WPs due to its half-metallic nature. Analysis with a low-energy effective Hamiltonian shows that this ideal Weyl semimetal is obtained in $EuCd_2As_2$ because the DPs are very close to the zone center and the ferromagnetic exchange splitting is large enough to push one pair of WPs to merge and annihilate at $\Gamma$ while the other pair survives. Furthermore, we predict that alloying with Ba at the Eu site can stabilize the ferromagnetic configuration and generate a single pair of Weyl points without application of a magnetic field.

# Introduction

Discoveries of symmetry-protected topological states in condensed matter physics and materials science have attracted great interest.[1-4] Among the different symmetries, maintaining time-reversal symmetry (TRS) $\Theta$ is key to protecting many topological states, such as topological insulator (TI), because $\Theta$ is antiunitary for electrons and maintains the degeneracy of Kramer pairs. Breaking TRS with magnetism can introduce other interesting topological states. For example, the quantum anomalous Hall effect (QAHE) has been realized in ferromagnetic TIs with magnetic dopants.[5] AFM-TIs have also been proposed[6] where an effective TRS can be defined by combining $\Theta$ and non-symmorphic translation $\tau_{1/2}$ ($S=\Theta\tau_{1/2}$). The preservation of $S$ can lead to topological $Z_2$ classification even though $\Theta$ is broken. The surface states of an AFM-TI can either gap out or form a surface Dirac point (DP), depending on whether the surface orientation breaks or preserves $S$, with the former being proposed to host topological axion states[7] with quantized magnetoelectric effects. Very recently, such an intrinsic AFM-TI has been reported in $MnBi_2Te_4$.[8-10]

TRS is also a key factor in topological semimetals where broken TRS can lift the degeneracy of bulk DPs into Weyl points separated in momentum space. Weyl points (WPs) were first proposed in noncollinear antiferromagnetic (AFM) pyrochlore iridates[11] and ferromagnetic (FM) $HgCr_2Se_4$[12], before they were realized in compounds with broken inversion symmetry (IS) $P$, such as TaAs[13-16] and $WTe_2$ families[17-21]. However, bulk DPs can also exist in AFM systems, such as CuMnAs, due to the protection of the combined $\Theta P$, although each of $\Theta$ and $P$ by itself is broken.[22] Clearly, the interplay of magnetism, crystalline symmetry, and topology provides a mechanism whereby magnetism can be used to control topological states.

For topological semimetals, a major challenge is to discover a material with a single pair of WPs in the full Brillouin zone (BZ) which would greatly simplify the elucidation of their unique transport and optical properties. $MnBi_2Te_4$ has recently been predicted to host a single pair of FM-WPs in the FM phase[8]. However, first-principles electronic structure calculations using experimentally determined lattice parameters found that the FM configuration of $MnBi_2Te_4$ is not semimetallic and is better regarded as an AFM-TI. The AFM ground state of $MnBi_2Te_4$ has a band gap of 200 meV and the FM configuration is 46 meV/f.u. less stable. To form a pair of WPs in $MnBi_2Te_4$, the insulating gap needs to be closed by expanding the structure in the basal plane or under negative pressure. Given such requirements, it is not straightforward to realize the single pair of FM-WPs in semiconducting $MnBi_2Te_4$. A pair of FM-WPs has also been predicted[12] to exist in $HgCr_2Se_4$ along the magnetization axis, but there is an additional nodal loop protected by mirror symmetry at $k_z=0$ plane.

Thus, a FM semimetal hosting the ideal case of a single pair of WPs has not been found yet. A better starting place to search for such an ideal semimetallic system is an AFM-DP semimetal or AFM-TI with a very small band gap.[23, 24] A recent study of the AFM semimetal $EuCd_2As_2$ found that a single pair of AFM-DP is protected by *SP* combination.[23] $EuCd_2As_2$ is AFM with a Neel temperature of 9.5 K[25, 26] and strong coupling is observed between transport and magnetism[27]. Resonant x-ray scattering measurements by Rahn *et al.*[24] have shown that the ground state magnetic structure of $EuCd_2As_2$ is an A-type AFM with moments aligned in the hexagonal layer. The A-type order with in-plane moments breaks the 3-fold rotational symmetry and turns AFM-DP into an AFM-TI state with a band gap of 10 meV. Additionally, Hua *et al.*[23] showed that each of the AFM-DPs can also be split into a pair of triply degenerate nodal points (TDNPs) when breaking *P*.

As shown in the calculation by Krishna et al.[28], the energy difference between AFM and FM is as small as ~2 meV/f.u, suggesting that the FM state is accessible under typical laboratory conditions. Rahn et al.[24] have shown that the saturation magnetic field to stabilize FM along c axis is about 1.5 T at 2 K. However, the topological properties of the FM configuration have not been studied yet. Here we use first-principles band structure calculations and discover that that a FM configuration with moments along c axis of EuCd$_2$As$_2$ actually hosts a single pair of WPs, a unique feature due to half metallic nature of the system. Furthermore, we propose that alloying with Ba at the Eu site introduces an effective negative pressure that can stabilize FM over AFM. By introducing exchange splitting terms explicitly in a 4-band low-energy Hamiltonian for the DPs, we find that a single pair of DPs will split into a single pair of WPs under the following conditions; (1) the DPs should be close to the BZ center and (2) the exchange splitting must be large enough to push two WPs from the opposite side of Γ point to merge and annihilate, while the other pair of WPs survive by being pushed away from the Γ point.

## Computational Methods

All density functional theory[29,30] (DFT) calculations with and without spin-orbit coupling (SOC) were performed with the PBE[31] exchange-correlation functional using a plane-wave basis set and projector augmented wave[32] method, as implemented in the Vienna Ab-initio Simulation Package[33,34] (VASP). Using maximally localized Wannier functions[35-37], tight-binding models were constructed to reproduce closely the band structure including SOC within ±1eV of the Fermi energy ($E_F$) with Eu *s-d-f*, Cd *s-p* and As *p* orbitals. The surface Fermi arcs and spectral functions were calculated with the surface Green's function methods[38-42]. In the DFT calculations, we used a kinetic energy cutoff of 318 eV, Γ-centered Monkhorst-Pack[43] (11×11×3) *k*-point mesh, and a Gaussian smearing of 0.05 eV. To account for the strongly localized Eu 4*f* orbitals, an onsite U=5.0 eV is used[23, 28]. For band structure calculations of EuCd$_2$As$_2$, the atomic basis positions and unit cell vectors are fixed to the experimental values[26]. To study the effect of alloying with Ba on the Eu site, a 20-atom supercell along $c$ with alternating Ba and Ba/Eu layers (at 50% Ba substitution in every other layer) is fully relaxed in the FM and AFM configurations. The magnitude of force on each atom is reduced to below 0.01 eV/Å.

## Results and Discussion

Figure 1 shows the bulk band structures and topological features of EuCd$_2$As$_2$ in different magnetic configurations. The crystal structure[26] of EuCd$_2$As$_2$, with the primitive hexagonal unit cell of 5 atoms in space group 164 (*P*–3*m1*), is shown in Fig.1(b). The Eu atoms form a simple hexagonal lattice at the 1*a* Wyckoff position. The As and Cd atoms at the 2*b* positions form the other four atomic layers with the sequence of -Cd-As-Eu-As-Cd- along the $c$ axis. The As layer binds to Eu on one side and Cd on the other side. The structure has inversion symmetry, *P*. The Eu half-filled 4*f* shell hosts a large magnetic moment of 7 $\mu_B$. In EuCd$_2$As$_2$, the Eu moments prefer an intralayer FM coupling and an interlayer AFM coupling along $c$ axis, i.e., an A-type AFM (AFMA), which requires doubling the unit cell along $c$ direction. The insets of Fig.1(c) and (d) show two such magnetic configurations by singling out Eu atoms with magnetic moment directions along $c$ (AFMAc) and along $a$ (AFMAa), respectively.

Although magnetism breaks $\Theta$, for the AFMA in EuCd$_2$As$_2$, $\Theta$ can be coupled with a lattice translation along $c$, giving the non-symmorphic TRS[6, 23] $S=\Theta\tau_{1/2}$. For example, starting with either AFMA configurations in the inset of Fig.1(c) or (d), the magnetic moments are switched to the opposite direction under $\Theta$ and a translation $\tau_{1/2}$ along $c$ returns the system to the starting configuration. The combination of *S* and *P* is antiunitary, $(SP)^2=-1$, thus the bands at each general k-point are doubly degenerate for AFMA EuCd$_2$As$_2$ in the presence of SOC due to Kramer's theorem.

For EuCd$_2$As$_2$, different magnetic configurations affect the topological features of the system. Figure 1 presents the band structures of different magnetic configurations including the ground state AFMAa configuration found experimentally[24]. Starting with the AFMAc band structure (Fig.1(a)), there is a sizable gap between valence and conduction bands along all the high-symmetric directions except near Γ. The zoom in along ΓA in Fig.1(c) shows the band inversion between Cd $s$ and As $p$ derived orbitals near Γ. For AFMAc, the crossing is not gapped out due to the protection from the 3-fold rotational symmetry similar to a class-I[44] DP in a non-magnetic system, resulting in a pair of AFM-DPs at (0.0, 0.0, ±0.004 Å$^{-1}$). When the magnetic moments are oriented along $a$ (Fig.1(d)) for AFMAa, the AFM-DPs are gapped out by breaking the 3-fold rotational symmetry and the system becomes an AFM-TI. Hua $et\ al.$[23] have also shown that the application of strain along the $a$ axis is another way to break the 3-fold rotational symmetry and open up a band gap in the AFMAc phase. The AFM-TI has a special gapped surface state if the open boundary is along $c$, resulting in half-integer quantum Hall effect and axion states. In summary, our DFT+U+SOC calculations and analysis agree with earlier studies[24, 28] for EuCd$_2$As$_2$ on the topological features of the AFMA states.

If the AFMAa ground state can be switched into a FMc configuration, then $S$ will be broken and the double degeneracy will be lifted. Rahn $et\ al.$[24] have shown that the saturation magnetic field to stabilize FM along $c$ axis is about 1.5 T at 2 K. In Fig.1(e), we show that yet another interesting topological feature, a single pair of WPs, will emerge in the FMc state. This is a unique feature due to the half-metallic nature of the system (see below). The Berry curvature calculation in Fig.1(f) indeed shows the source and sink for this single pair of WPs at (0.0, 0.0, ±0.03 1/Å) on either side of the Γ point. Note that the distance between FM-WPs is increased compared to AFM-DPs because the band crossing is pushed away from Γ.

When projected onto the crystalline (001) surface of EuCd$_2$As$_2$, the WPs of opposite chirality overlap in the FMc state, thus there will be no open Fermi arcs. In contrast, the single pair of WPs do not overlap when projected on (110) surface and open Fermi arcs connecting the projected WPs are expected. Figure 2 shows such open Fermi arcs on both the top and bottom (110) surfaces of EuCd$_2$As$_2$ with Cd termination. In Fig.2(a) and (b) the Fermi arcs connect two hole pockets right at $E_F$, each on the opposite (110) surface. At higher energies, the hole pockets shrink and eventually reach the energy of the WPs at $E_F$+48 meV where the size of the Fermi arcs is maximal, spanning 0.06 1/Å. This situation corresponds to the ideal case of two open Fermi arcs connecting a single pair of WPs in the full BZ, each on the opposite surface, which has only been described schematically and without confirmation in any material system. Here we predict that such an ideal case of a single pair of WPs can be realized in FM EuCd$_2$As$_2$.

Normally, we expect that one pair of DPs will split into two pairs of WPs when breaking $\Theta$ or, in our case, $S$. Where does the other pair of WPs go in EuCd$_2$As$_2$? To clarify why and how a single pair of FM-WPs can arise from a single pair of AFM-DPs in EuCd$_2$As$_2$ upon breaking $S$, we present the evolution of band structures along ΓA direction in Fig.3 for different magnetic configurations. First, we turn off the magnetism in EuCd$_2$As$_2$ by treating the Eu 4$f$ orbitals as core electrons. The system retains both $\Theta$ and $P$ symmetries. The non-magnetic band structure without SOC is shown in Fig.3(a). It is a semiconductor with a band gap of 200 meV. The valence band has orbital degeneracy of As $p_x$ and $p_y$, thus is 4-fold degenerate when spin degeneracy is included. The conduction band is derived from Cd $s$ orbital and has 2-fold degeneracy from spin. Turning on SOC in Fig.3(b), the orbital degeneracy between As $p_x$ and $p_y$ is lifted, but still maintains 2-fold degeneracy from spin because of $\Theta$ and $P$. The band gap is reduced to 100 meV.

For AFMA with 4$f$ moment on Eu atoms and the unit cell doubled along $c$ axis, $\Theta$ is broken, but $S$ is not broken. The combination of $S$ and $P$ still gives 2-fold degeneracy for a general $k$-point. Comparing Fig.3(c) to Fig.3(a) (without SOC) and Fig.3(d) to Fig.3(b) (with SOC), the main effect of AFMAc order is a band folding due to doubling of the unit cell along $c$. Without SOC in Fig.3(c), the valence band has 4-fold degeneracy from both the spin and orbital degeneracy (As $p_x$ and $p_y$), while the conduction band derived from Cd $s$ orbital only has 2-fold spin degeneracy. The orbital degeneracy for the valence band can be clearly seen in the color-coding of the folded branch from −0.7 at Γ to −0.5 eV at A, which has the opposite dispersion to the valence band. The lower branch of the valence band crosses another band mostly of As $p_z$ (Eu $f$) character near Γ (A). The conduction band has flat folded branch with Cd $s$ character and the higher energy conduction band is mostly of Cd $s$, $p_z$ and As $p_z$ character. Each of these three conduction bands has 2-fold spin degeneracy.

Without SOC, the AFMA valence and conduction bands still do not cross, and a small band gap remains at Γ. When SOC is turned on in Fig.3(d), the orbital degeneracy of As $p_x$ and $p_y$ valence bands are lifted. Because of SOC, now $j$ is good quantum number. The valence band with a total angular momentum of $J=3/2$ is split into two, one with $j_z=\pm 3/2$ and the other $j_z=\pm 1/2$; each is 2-fold degenerate due to $S$ and $P$. The $J=1/2$ conduction band is also 2-fold degenerate. Importantly, SOC induces a band inversion around Γ between the valence $\left|\frac{3}{2},\pm\frac{1}{2}\right\rangle$ and conduction $\left|\frac{1}{2},\pm\frac{1}{2}\right\rangle$. The system becomes a topological semimetal with a single pair of AFM-DPs, as discussed by Hua *et al*.[23]

We now consider the case for FM order, where $S$ is broken and the 2-fold spin degeneracy is lifted. Without SOC, comparing Fig.3(e) to Fig.3(a), there is a large exchange splitting effect, giving two sets of bands, spin up and down. This results in a unique half-metallic feature where only the spin-up valence and conduction bands cross each other at $E_F$, while the spin-down bands do not cross and maintain a gap. This crossing between the spin-up valence and conduction bands is a TDNP because the valence band has the 2-fold orbital degeneracy between As $p_x$ and $p_y$. When SOC is turned on for the FMc configuration in Fig.3(f), both the orbital and spin degeneracies are fully lifted, and the 4-fold degenerate valence band in Fig.1(a) now becomes four individual bands. The spin and orbital degeneracies of other bands (including the conduction band) are already fully lifted by the broken $S$ in the FM state. The crossing of half-metallic spin-up valence and conduction bands near $\Gamma$ is preserved with SOC. Importantly, the degeneracy of this crossing is reduced from three for TDNP to two, thus generating a WP.

The connection between the AFM-DPs (Fig.3(d)) and the FM-WPs (Fig.3(f)) can be understood by unfolding the bands and then lifting the spin degeneracy due to the breakdown of $(SP)^2=-1$. The usual scenario of a single DP splitting into two WPs does not work here because of the half-metallicity. Or equivalently, the WP closer to $\Gamma$ can merge with another WP of the opposite chirality from the other side of $\Gamma$ and annihilate to form a gap.

To get a more generic picture of splitting a single pair of DPs into a single pair of WPs in a half-metal, we introduce the relevant low-energy Hamiltonian. We start with the 4-band $k.p$ model for AFM-DP in EuCd$_2$As$_2$ with $\left|\frac{1}{2},+\frac{1}{2}\right\rangle, \left|\frac{3}{2},+\frac{3}{2}\right\rangle$, $\left|\frac{1}{2},-\frac{1}{2}\right\rangle$ and $\left|\frac{3}{2},-\frac{3}{2}\right\rangle$ in that order. This is similar to class-I DP in Na$_3$Bi[45] and Cd$_3$As$_2$[46], except that the off-diagonal matrix elements are in the $k_\pm$ order. We can add the FM exchange splitting as two terms, one is orbital dependent $h_1\sigma_z\otimes\tau_z$ and the other spin dependent $h_2\sigma_z \otimes I$, where $h_1$ and $h_2$ are exchange splitting parameters, $\sigma$ and $\tau$ are the Pauli matrices for spin and orbital, respectively, and $I$ is the identity matrix. Then the total Hamiltonian is

$$H_{eff}(k) = \varepsilon_0(k) +$$

$$\begin{pmatrix} M(k) + h_1 + h_2 & Ak_+ & 0 & Bk_+ \\ Ak_- & -M(k) - h_1 + h_2 & Bk_+ & 0 \\ 0 & Bk_- & M(k) - h_1 - h_2 & -Ak_- \\ Bk_- & 0 & -Ak_+ & -M(k) + h_1 - h_2 \end{pmatrix} \quad (1)$$

where $k_\pm = k_x \pm ik_y$, $M(k) = M_0 - M_1 k_z^2 - M_2(k_x^2 + k_y^2)$ and $\varepsilon_0(k) = C_0 + C_1 k_z^2 + C_2(k_x^2 + k_y^2)$. After diagonalization, the solutions for the eigenvalues are

$$E(k) = \varepsilon_0(k) \pm \sqrt{(M(k) + sgn\, h_1)^2 + (A^2 + B^2)(k_x^2 + k_y^2)} + sgn\, h_2 \quad (2)$$

where *sgn* takes on ± independently from the one before the square root. When fitted to DFT calculated band structures of EuCd$_2$As$_2$, the following parameters are obtained, $C_0$=–0.0005 eV, $C_1$=47.8237 eVÅ$^2$, $C_2$=55.0803 eVÅ$^2$, $M_0$=–0.0039 eV, $M_1$=–54.5293 eVÅ$^2$, $M_2$=1.6425 eVÅ$^2$ and $\tilde{A}$=2.6902 eVÅ for $\tilde{A}^2 = A^2 + B^2$.

In Fig.4, we plot the bands from the low-energy effective Hamiltonian (Eqn.(2)) with different values of $h_1$ and $h_2$. When $h_1 = h_2 = 0$ in Fig.4(a), we have a 4-fold degenerate AFM-DP due to the combined *SP*. When *S* is broken with $h_1$=3.0 meV and $h_2$=0 in Fig.4(b), a pair of FM-WPs split from the AFM-DPs. By increasing $h_2$ such that $h_1 = h_2 = 3.0$ meV in Fig.4(c), two TDNPs are produced because the valence band is 2-fold degenerate, but not the conduction band. An increase of $h_1$ to 6.0 meV in Fig.4(d) is enough to create a gap between one pair of valence and conduction bands. In effect, one of the two WPs pairs on opposite sides of Γ meet and annihilate, while the other pair of WPs is pushed further away from Γ. This can happen in EuCd$_2$As$_2$ is because AFM-DPs are very close to Γ at ±0.004 (1/Å), comparing to ±0.084 and ±0.032 (1/Å) in Na$_3$Bi[45] and Cd$_3$As$_2$[46], respectively.

Although the FM configuration of EuCd$_2$As$_2$ hosts the desirable single pair of WPs, experiment has shown that the ground state of EuCd$_2$As$_2$ is AFMA with magnetic moments lying in-plane. To stabilize FM vs. AFMA, one can apply a saturation magnetic field of about 1.5 T along the *c* axis at 2 K as shown by Rahn *et al*.[24] Due to the small difference in energy between AFMA and FM states, another possibility is to stabilize the FM state by expanding the lattice using strain or by the application of an effective negative pressure by alloying. For example, in the same crystal structure, the lattice constants of BaCd$_2$As$_2$ are ~5% larger than EuCd$_2$As$_2$. As a proof of principle, we have constructed a 20-atom supercell along *c* with an alternating Ba and Eu layers (i.e. with equal concentrations of Ba and Eu). For pure EuCd$_2$As$_2$, the DFT+U+SOC relaxed lattice constants of *a*=4.49 Å and *c*=7.38 Å agree very well with experimental data[26] of *a*=4.4499 Å and *c*=7.35 Å (within 1%). For EuBaCd$_4$As$_4$, the lattice constants increase to *a*=4.53 Å and *c*=7.56 Å, and FMc is 0.25 meV/f.u. more stable than AFMAc. Figure 5 shows the band structures of EuBaCd$_4$As$_4$ along ΓA for the AFMAc and FMc configurations. The single pair of DPs for AFMAc in Fig.5(a) and single pair of WPs for FM in Fig.5(b) are retained. The Ba-derived bands do not affect the half-metallic feature and the crossings between As *p* and Cd *s* derived bands near the $E_F$. The increased lattice constants obtained from alloying with Ba at the Eu site result in a ground state FMc configuration with a single pair of WPs.

## Conclusion

In conclusion, first-principles band structure calculations show that EuCd$_2$As$_2$ in a ferromagnetic configuration can host an ideal case of a single pair of Weyl points due to the half-metallic nature of the system. Analysis with a low-energy effective Hamiltonian shows that this ideal case is obtained in $EuCd_2As_2$ because the Dirac points in the AFMA state are very close to the zone center and the ferromagnetic exchange splitting is large enough to push one pair of WPs, one from each side of $\Gamma$, to merge and annihilate. Furthermore, we have shown that by alloying with Ba at the Eu site, the ferromagnetic configuration can become the ground state. $EuCd_2As_2$ thin films grown with molecular beam epitaxy along a (110) surface and stabilized into the FM state with strain engineering or Ba alloying will be desirable systems to investigate such an ideal case of a single pair of Weyl points.

**ACKNOWLEDGMENT**

This work was supported as part of the Center for the Advancement of Topological Semimetals, an Energy Frontier Research Center funded by the U.S. Department of Energy Office of Science, Office of Basic Energy Sciences through the Ames Laboratory under its Contract No. DE-AC02-07CH11358.

\* llw@ameslab.gov

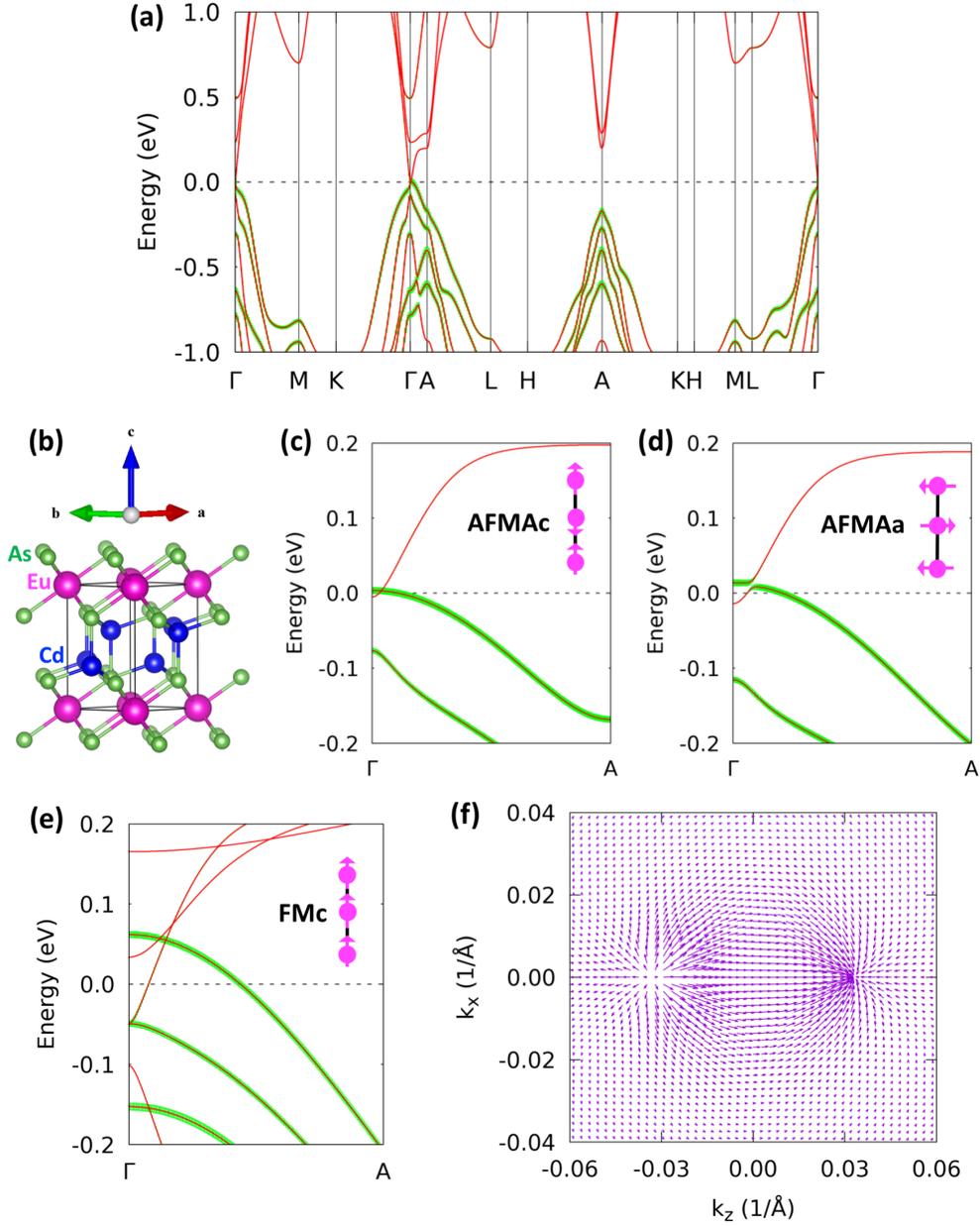

Figure 1. (a) Band structure of $EuCd_2As_2$ in PBE+U+SOC for A-type AFM with magnetic moment along $c$ axis (AFMAc). (b) crystal structure of $EuCd_2As_2$ in space group 164 (*P–3m1*). Band structure along ΓA for (c) AFMAc, (d) moments in-plane along major axis (AFMAa) breaking 3-fold rotational symmetry, and (e) FM moment along c (FMc) breaking the combined $S=\Theta\tau_{1/2}$ symmetry. The green shade over the band structure stands for the projection of As *p* orbitals. (f) Berry curvature on the $k_x$-$k_z$ plane showing chirality for the single pair of Weyl points in (e).

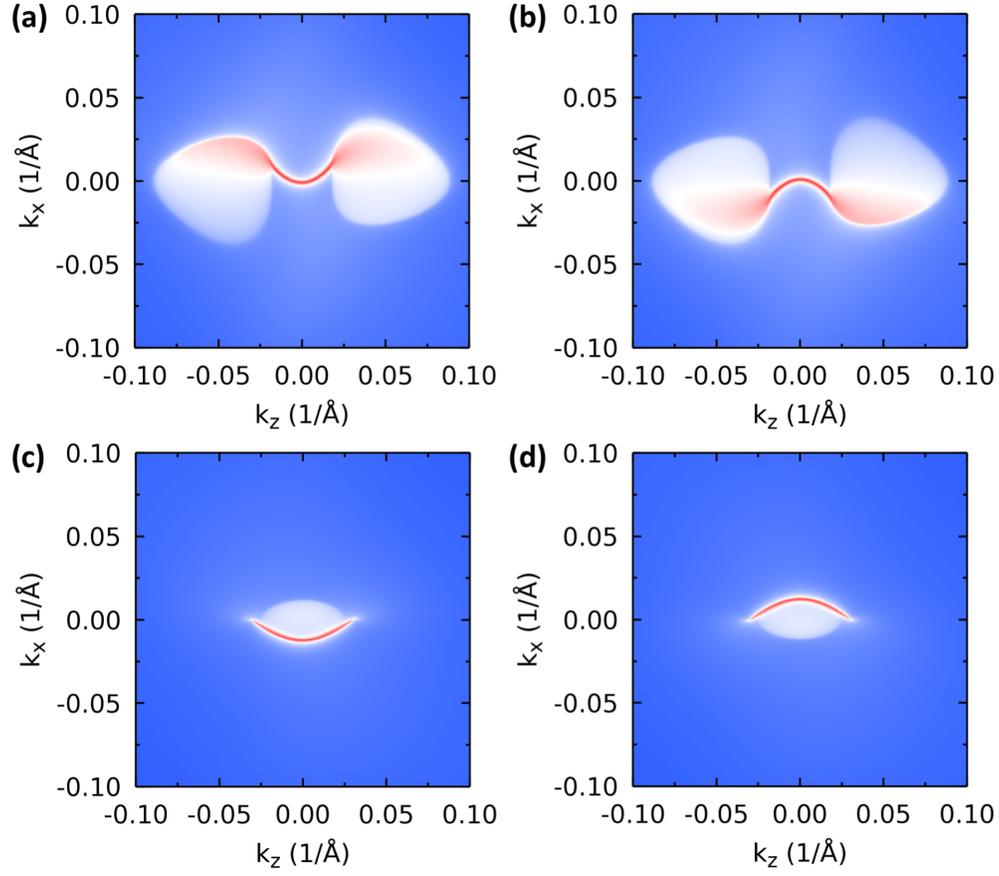

Figure 2. Fermi arcs at $E_F$ on the (a) top and (b) bottom (110) surfaces of $EuCd_2As_2$ in FMc with a single pair of Weyl points; and those at $E_F+0.048$ eV on the (c) top and (d) bottom (110) surfaces. The color scheme of the intensity from high to low is red, white and blue.

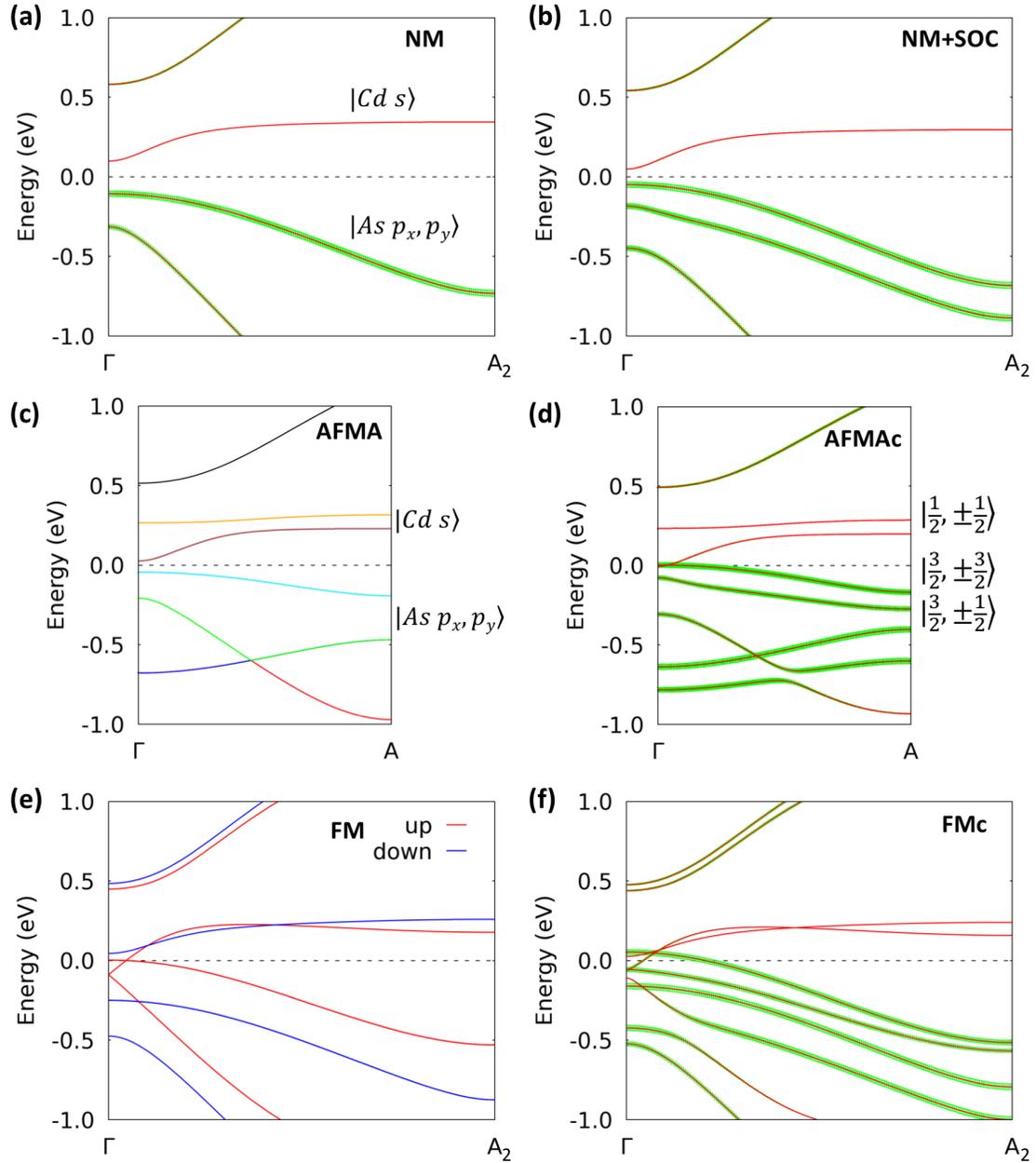

Figure 3. Bulk band structure of $EuCd_2As_2$ along ΓA direction for non-magnetic (NM) without Eu 4*f* in (a) PBE and (b) PBE+SOC; AFM in (c) PBE+U and (d) PBE+U+SOC; and FM in (e) PBE+U and (f) PBE+U+SOC. The NM and FM have the primitive 5-atom unit cell. The AFM has the 10-atom unit cell by doubling along *c* direction. The green shade shows the projection on As *p* orbitals. Each band in (c) has a different color to show the orbital degeneracy in the valence band.

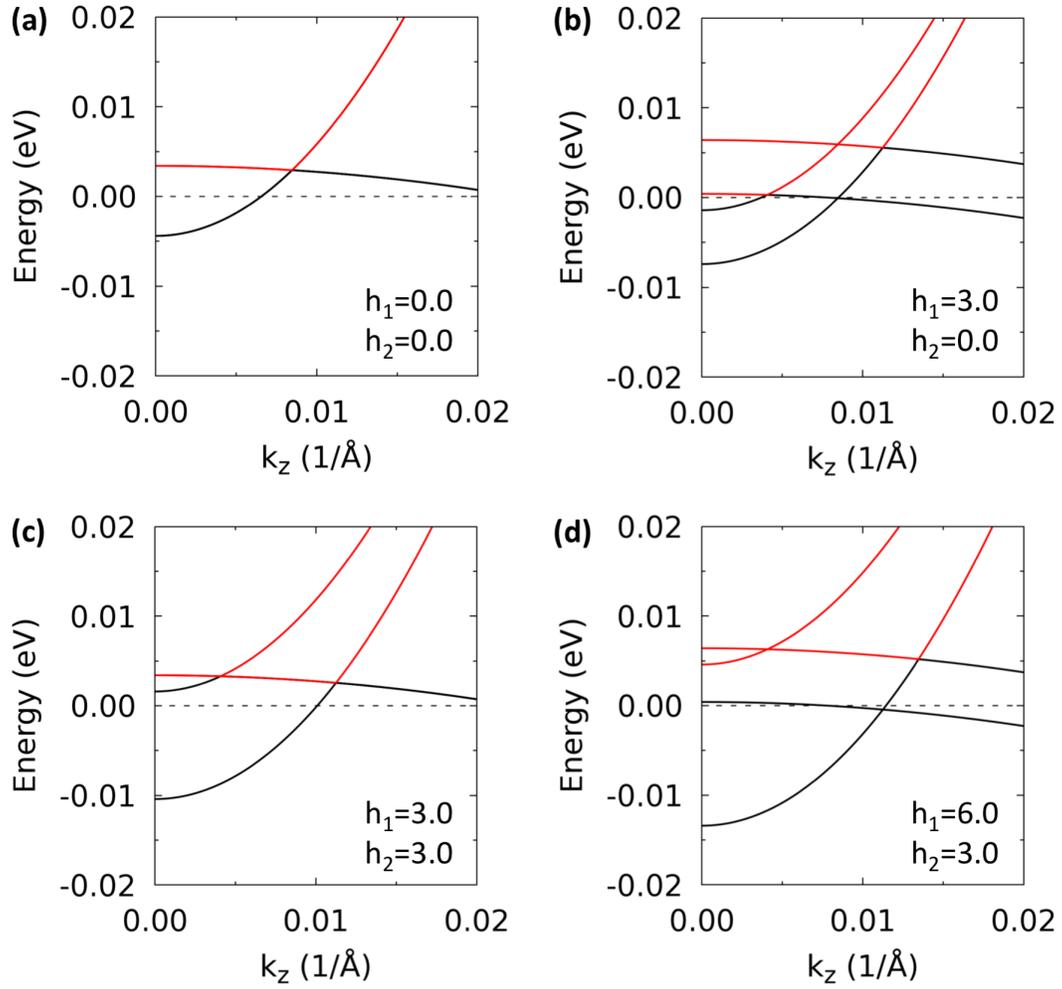

Figure 4. Band structure near $\Gamma$ along $\Gamma A$ of the 4-band low-energy Hamiltonian based on EuCd$_2$As$_2$ with different values of exchange splitting parameters $h_1$ and $h_2$ in meV.

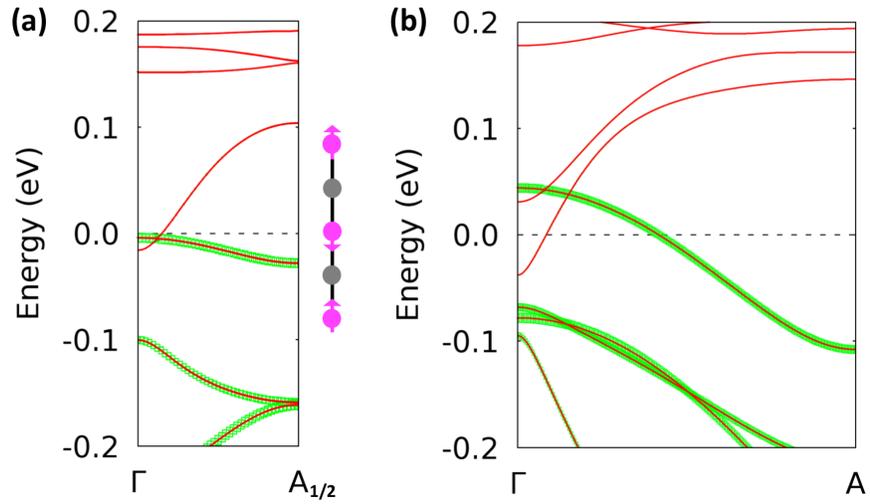

Figure 5. Band structure of 50% Ba (gray) substituted $EuCd_2As_2$, or $EuBaCd_4As_4$, for (a) AFMAc and (b) FMc with moment along *c*. Unit cells are fully relaxed and FMc is preferred. The green shade shows the projection on As *p* orbitals.

# References

1. M. Z. Hasan, C. L. Kane, Colloquium: Topological insulators. *Rev Mod Phys* **82**, 3045 (2010).

2. X.-L. Qi, S.-C. Zhang, Topological insulators and superconductors. *Rev Mod Phys* **83**, 1057-1110 (2011).

3. C. K. Chiu, J. C. Y. Teo, A. P. Schnyder, S. Ryu, Classification of topological quantum matter with symmetries. *Rev Mod Phys* **88**, 035005 (2016).

4. N. P. Armitage, E. J. Mele, A. Vishwanath, Weyl and Dirac semimetals in three-dimensional solids. *Rev Mod Phys* **90**, 015001 (2018).

5. C. Z. Chang *et al.*, Experimental Observation of the Quantum Anomalous Hall Effect in a Magnetic Topological Insulator. *Science* **340**, 167-170 (2013).

6. R. S. K. Mong, A. M. Essin, J. E. Moore, Antiferromagnetic topological insulators. *Phys Rev B* **81**, 245209 (2010).

7. A. M. Essin, J. E. Moore, D. Vanderbilt, Magnetoelectric Polarizability and Axion Electrodynamics in Crystalline Insulators (vol 102, 146805, 2009). *Phys Rev Lett* **103**, 146805 (2009).

8. D. Zhang *et al.*, Topological axion states in magnetic insulator MnBi$_2$Te$_4$ with the quantized magnetoelectric effect. arXiv:1808.08014 (2018).

9. Y. Gong *et al.*, Experimental realization of an intrinsic magnetic topological insulator. arXiv:1809.07926 (2018).

10. M. M. Otrokov *et al.*, Prediction and observation of the first antiferromagnetic topological insulator. arXiv:1809.07389 (2018).

11. X. G. Wan, A. M. Turner, A. Vishwanath, S. Y. Savrasov, Topological semimetal and Fermi-arc surface states in the electronic structure of pyrochlore iridates. *Phys Rev B* **83**, 205101 (2011).

12. G. Xu, H. M. Weng, Z. J. Wang, X. Dai, Z. Fang, Chern Semimetal and the Quantized Anomalous Hall Effect in HgCr2Se4. *Phys Rev Lett* **107**, 186806 (2011).

13. S. Y. Xu *et al.*, Discovery of a Weyl fermion semimetal and topological Fermi arcs. *Science* **349**, 613-617 (2015).

14. B. Q. Lv *et al.*, Experimental Discovery of Weyl Semimetal TaAs. *Phys Rev X* **5**, 031013 (2015).

15. H. M. Weng, C. Fang, Z. Fang, B. A. Bernevig, X. Dai, Weyl Semimetal Phase in Noncentrosymmetric Transition-Metal Monophosphides. *Phys Rev X* **5**, 011029 (2015).

16. S. M. Huang *et al.*, A Weyl Fermion semimetal with surface Fermi arcs in the transition metal monopnictide TaAs class. *Nat Commun* **6**, 7373 (2015).

17. L. N. Huang *et al.*, Spectroscopic evidence for a type II Weyl semimetallic state in MoTe2. *Nat Mater* **15**, 1155-1160 (2016).

18. A. A. Soluyanov *et al.*, Type-II Weyl semimetals. *Nature* **527**, 495-498 (2015).

19. Y. Wu *et al.*, Temperature-Induced Lifshitz Transition in WTe2. *Phys Rev Lett* **115**, 166602 (2015).

20. I. Belopolski *et al.*, Fermi arc electronic structure and Chern numbers in the type-II Weyl semimetal candidate MoxW1-xTe2. *Phys Rev B* **94**, 085127 (2016).

21. Y. Wu *et al.*, Observation of Fermi arcs in the type-II Weyl semimetal candidate WTe2. *Phys Rev B* **94**, 121113(R) (2016).

22. P. Z. Tang, Q. Zhou, G. Xu, S. C. Zhang, Dirac fermions in an antiferromagnetic semimetal. *Nat Phys* **12**, 1100-1104 (2016).

23. G. Hua *et al.*, Dirac semimetal in type-IV magnetic space groups. *Phys Rev B* **98**, 201116 (2018).

24. M. C. Rahn *et al.*, Coupling of magnetic order and charge transport in the candidate Dirac semimetal $\mathrm{EuCd}_{2}\mathrm{As}_{2}$. *Phys Rev B* **97**, 214422 (2018).

25. A. Artmann, A. Mewis, M. Roepke, G. Michels, AM(2)X(2) compounds with the CaAl2Si2-Type structure .11. Structure and properties of ACd(2)X(2) (A: Eu, Yb; X: P, As, Sb). *Z Anorg Allg Chem* **622**, 679-682 (1996).

26. I. Schellenberg, U. Pfannenschmidt, M. Eul, C. Schwickert, R. Pottgen, A Sb-121 and Eu-151 Mossbauer Spectroscopic Investigation of EuCd2X2 (X = P, As, Sb) and YbCd2Sb2. *Z Anorg Allg Chem* **637**, 1863-1870 (2011).

27. H. P. Wang, D. S. Wu, Y. G. Shi, N. L. Wang, Anisotropic transport and optical spectroscopy study on antiferromagnetic triangular lattice EuCd2As2: An interplay between magnetism and charge transport properties. *Phys Rev B* **94**, 045112 (2016).

28. J. Krishna, T. Nautiyal, T. Maitra, First-principles study of electronic structure, transport, and optical properties of $\mathrm{EuCd}_{2}\mathrm{As}_{2}$. *Phys Rev B* **98**, 125110 (2018).

29. P. Hohenberg, W. Kohn, Inhomogeneous Electron Gas. *Phys. Rev.* **136**, B864-B871 (1964).

30. W. Kohn, L. J. Sham, Self-Consistent Equations Including Exchange and Correlation Effects. *Phys. Rev.* **140**, A1133-A1138 (1965).

31. J. P. Perdew, K. Burke, M. Ernzerhof, Generalized gradient approximation made simple. *Phys Rev Lett* **77**, 3865-3868 (1996).

32. P. E. Blöchl, Projector Augmented-Wave Method. *Phys Rev B* **50**, 17953-17979 (1994).

33. G. Kresse, J. Furthmuller, Efficient Iterative Schemes for Ab initio Total-Energy Calculations Using a Plane-Wave Basis Set. *Phys Rev B* **54**, 11169-11186 (1996).

34. G. Kresse, J. Furthmuller, Efficiency of Ab-initio Total Energy Calculations for Metals and Semiconductors Using a Plane-Wave Basis Set. *Comp Mater Sci* **6**, 15-50 (1996).

35. N. Marzari, D. Vanderbilt, Maximally localized generalized Wannier functions for composite energy bands. *Phys Rev B* **56**, 12847-12865 (1997).

36. I. Souza, N. Marzari, D. Vanderbilt, Maximally localized Wannier functions for entangled energy bands. *Phys Rev B* **65**, 035109 (2001).

37. N. Marzari, A. A. Mostofi, J. R. Yates, I. Souza, D. Vanderbilt, Maximally localized Wannier functions: Theory and applications. *Rev Mod Phys* **84**, 1419-1475 (2012).

38. Q. Wu, S. Zhang, H.-F. Song, M. Troyer, A. A. Soluyanov, WannierTools: An open-source software package for novel topological materials. *Computer Physics Communications* **224**, 405-416 (2018).

39. D. H. Lee, J. D. Joannopoulos, Simple Scheme for Surface-Band Calculations .1. *Phys Rev B* **23**, 4988-4996 (1981).

40. D. H. Lee, J. D. Joannopoulos, Simple Scheme for Surface-Band Calculations .2. The Greens-Function. *Phys Rev B* **23**, 4997-5004 (1981).

41. M. P. L. Sancho, J. M. L. Sancho, J. Rubio, Quick Iterative Scheme for the Calculation of Transfer-Matrices - Application to Mo(100). *J Phys F Met Phys* **14**, 1205-1215 (1984).

42. M. P. L. Sancho, J. M. L. Sancho, J. Rubio, Highly Convergent Schemes for the Calculation of Bulk and Surface Green-Functions. *J Phys F Met Phys* **15**, 851-858 (1985).

43. H. J. Monkhorst, J. D. Pack, Special Points for Brillouin-Zone Integrations. *Phys Rev B* **13**, 5188-5192 (1976).

44. B. J. Yang, N. Nagaosa, Classification of stable three-dimensional Dirac semimetals with nontrivial topology. *Nat Commun* **5**, 4898 (2014).

45. Z. J. Wang *et al.*, Dirac semimetal and topological phase transitions in A(3)Bi (A = Na, K, Rb). *Phys Rev B* **85**, 195320 (2012).

46. Z. J. Wang, H. M. Weng, Q. S. Wu, X. Dai, Z. Fang, Three-dimensional Dirac semimetal and quantum transport in Cd3As2. *Phys Rev B* **88**, 125427 (2013).